\documentclass[aps,nofootinbib,floats,superscriptaddress]{revtex4}

\usepackage{graphicx}
\usepackage{epstopdf}
\usepackage{amssymb}
\usepackage{amsmath,bm}
\usepackage{psfrag}
\usepackage{epsfig}
\usepackage{float}
\usepackage{bm}
\usepackage{textcomp}

\begin{document}

\title{Effect of Electromigration on Onset of Morphological Instability of a Nanowire}

\author{Mikhail Khenner\footnote{Corresponding
author. E-mail: mikhail.khenner@wku.edu.}}
\affiliation{Department of Mathematics, Western Kentucky University, Bowling Green, KY 42101, USA}
\affiliation{Applied Physics Institute, Western Kentucky University, Bowling Green, KY 42101, USA}

\begin{abstract}
\noindent

Solid cylindrical nanowires are vulnerable to a Rayleigh-Plateau-type morphological instability. The instability results in a wire breakup, followed  
by formation of a chain array of spherical nanoparticles.  In this paper, a base model of a morphological instability of a 
nanowire on a substrate in the applied electric field directed along a nanowire axis is considered.
Exact analytical solution is obtained for $90^\circ$ contact angle and, assuming axisymmetric perturbations, for a free-standing wire. 
The latter solution extends the 1965 result by Nichols and Mullins without electromigration effect 
(F.A. Nichols and W.W. Mullins, ``Surface-(Interface-) and volume-diffusion contributions to morphological changes driven by capillarity", 
\textit{Trans. Metall. Soc. AIME}$\;$ {\bf 233},  1840–1848 (1965)).
For general contact angles the neutral stability is determined numerically. It is shown that a stronger applied electric field (a stronger current)
results in a larger instability growth rate and a decrease of the most dangerous unstable wavelength; 
in experiment, the latter is expected to yield more dense chain array of nanoparticles. 
Also it is noted that a wire crystallographic orientation on 
a substrate has larger impact on stability in a stronger electric field and that a simple switching of the polarity of electrical contacts, i.e. the reversal of the direction of the applied electric field,
may suppress the instability development and thus a wire breakup would be prevented. 
A critical value of the electric field that is required for such wire stabilization is obtained.

\end{abstract}

\date{\today}
\maketitle


\section{Introduction}
\label{Intro}

Theoretical studies of morphological stability and evolution of single-crystal solid wires (aka cylinders, rods, or whiskers) by surface diffusion have long history that begins with 
the 1965 work by Nichols and Mullins \cite{NM2,NM1}. These authors considered linear stability of a free-standing wire with respect to axisymmetric perturbations 
and found \cite{NM2} the dimensionless perturbations growth rate $\sigma=k^2(1-k^2)$, which is a classical expression of the Rayleigh-Plateau instability. In dimensional coordinates this implies that an axisymmetric perturbation induces instability if and only if
the wavelength is longer than the circumference of the undisturbed cylinder. More studies followed, where instability was investigated in various 
increasingly complicated settings, such as surface energy anisotropy, stress, and contact lines with the substrate \cite{C,McCVoorhees,BBW,WMVD,KDMV,GM,GMM,MyJCP}. 
These studies provided better understanding of the mechanisms responsible for wire breakup \cite{KTBECKN,KTEBCKN,LBPDVC,KT,XLL}. 
Recent models and computations are focused on solid-state dewetting of islands of various shapes, including wires \cite{BTS,WZSB,JWZ,BRTMP,EWCWGBCCT}. 

Electromigration \cite{Huntington,HoKwok,REW,SK}, on the other hand, has been modeled extensively as a simple and inexpensive technique to bias surface diffusion and 
thus affect and shape morphological and compositional instabilities and evolutions \cite{OPLE,KKHV,DDF,QM,SV,K,SGM,SGM1,DM,CMECPL,MySurfSci,MyPhysRevMat,SMSN,LECMC,CMCL}. 
Understanding surface electromigration-driven 
instabilities of wires is important for reliable and scalable manufacturing of nanocontacts \cite{PLAPM,VFDMSKBM,AGLCH}.  
Applied multi-physics models of nanowire failure due to electromigration were published \cite{BUS},
however, to our knowledge, there is no publications that theoretically address the fundamentals of electromigration-driven wire instability and evolution. 

In 1996, McCallum \textit{et al.} \cite{McCVoorhees} published the influential paper on stability of a wire with isotropic surface energy. 
In their work, the wire is assumed deposited onto a substrate, with the wire surface making two contact lines with the latter.
The very special feature of their 
analysis is that they allowed arbitrary contact angles with the substrate and did not assume the axial symmetry of perturbations. As the particular case, they 
determined that $\sigma(k)$ for a free-standing wire coincides with $\sigma(k)$ from Nichols and Mullins' paper \cite{NM2}.
These authors also found that the substrate stabilizes the wire, with the wire being most stable for small contact angles. 
Note that the model \cite{McCVoorhees} is still the base model, 
since it does not include any of the multitude of physical effects that typically influence morphological stability and evolution at the nanoscale. 
Some of these effects are the anisotropy of a surface energy (that often results in faceting \cite{GM,WZSB,JWZ,Korzec,OL}), 
stresses and strains \cite{BRTMP},
substrate wetting by the film \cite{K,TPL,OPL,KTL,K2008}, 
quantum size effect \cite{OPL,MSMSE_Khenner}, and electromigration.

The purpose of this paper is to re-visit 
the analysis in Ref. \cite{McCVoorhees} for the situation of electromigration bias of surface diffusion. In the next section we set up the 
base electromigration model, which is followed by the analysis in sections \ref{Res1}-\ref{Res3}.

\emph{Remark 1.}$\;$ Our physico-mathematical model is firmly rooted in the classical analysis of morphological evolution of solid surfaces 
by surface diffusion of adsorbed atoms, i.e. the mobile atoms in the surface layer, as was pioneered by Mullins \cite{Mullins95} 
and Cahn and co-workers \cite{CT94}.  
In the decades since the model was introduced, it has proved its usefulness and validity countless times, as is evidenced by the publications  
cited above. 
This author believes that it makes little sense to pile up the physical effects until the primary effect of interest (electromigration) is well understood in the context 
of morphological stability, evolution, and breakup of nanowires. This paper is the first step in that direction. Notice that in a Mullins-type model an 
effect is introduced via a term in a chemical potential of adatoms, i.e. effects are additive \cite{Mullins95,MSMSE_Khenner}.
Additional effects may be introduced and analyzed in the future, including some mentioned above and a more complex ones, such as thermal runaway - 
a harmful side effect of Joule heating that may result in a wire meltdown as its radius at the point of rupture tends to zero \cite{T}.    
Regarding electromigration, a classical phenomenon in solid state physics, 
inexplicably, its effect on morphological stability and evolution of wires was not considered theoretically or computationally.
This gap in the knowledge
is particularly striking, given that applied physics and phenomenology of electromigration-driven breakup of nanowires has been a fairly active area of experiment research 
in the community of nanoelectronics and device reliability engineers since at least 2004 (the year of the earliest publication this author was able to find; 
see Ref. \cite{H} for comprehensive review of the progress recently made).

\emph{Remark 2.}$\;$ Studies of charged liquid jets subjected to applied electric fields proliferate in the theoretical fluid mechanics community, 
see e.g. Ref. \cite{XYQF} and the references therein. In Ref. \cite{XYQF} it is shown that the axial
electric field will promote the predominance of asymmetric (i.e., non-axisymmetric) instability over the axisymmetric mode, 
which causes the bending motion in most experimental observations. Whether similar situation occurs for solid wires subjected to surface electromigration 
remains to be seen in a future computational study of morphological evolution \cite{BTS,WZSB,JWZ,EWCWGBCCT}. Without electromigration, it was shown that 
for free-standing wires the axisymmetric 
instability is dominant \cite{BBW} (at least when the surface energy is isotropic).    

\section{The model}
\label{Model}

In this section the base model of wire morphological evolution in the applied electric field is set up, starting from the geometry of a wire on a substrate. 
Surface energy anisotropy, bulk and contact line stresses, wetting, and other factors complicating analysis are ignored.
Apart from introducing surface electromigration via the term $\bm{\nabla}_s\cdot\left[M(\phi) E\cos{(\phi)}\bm{e}_y\right]$ \cite{SK,SGM,DM} in Eq. (\ref{Vn})
and the corresponding modification in the boundary condition (\ref{bc3}), the mathematical statement of the model is identical to Ref. \cite{McCVoorhees} 
(see below for the description of all mathematical symbols and physical quantities seen here).
The notations follow closely Refs. \cite{McCVoorhees,SK}. 

A mere local approximation for the surface electric field will be assumed, since it yields the correct result for the 
real part, $\sigma_r$ of the rate of growth or decay of the perturbation from the initial surface morphology \cite{SGM1}. $\sigma_r$ completely 
characterizes the film instability, which is the goal of this paper. Non-local electric field via the solution of the boundary-value problem for the 
electric potential leads to a non-zero imaginary part of $\sigma$ \cite{SK,SGM,B1}, i.e. a waves traveling on the surface, emerge. 
Analysis of traveling waves is deferred to future work.

It will be seen that the base model admits a semi-analytical solution for neutral stability only, that is, in contrast to Ref. \cite{McCVoorhees}, 
analytical determination of the 
instability growth rate (the dispersion relation) is 
analytically impossible even for this base model. Valuable insights are gained from the analytical solution of neutral stability. 
It is expected that accounting for any additional physical effect will require a fully numerical solution of a (linear) stability problem (and, of course, 
of a nonlinear morphology evolution problem).

\begin{figure}[H]
\vspace{-0.2cm}
\centering
\includegraphics[width=2.5in]{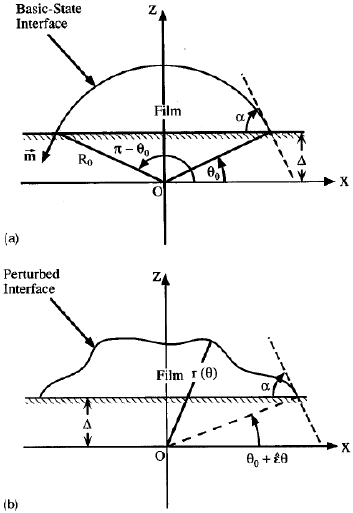}
\vspace{-0.15cm}
\caption{Cross section of a wire. (a) an equilibrium shape, (b) a perturbed state. The cylindrical coordinate system used to describe a wire is shown. 
The $y$ axis is perpendicular to the $x$ and $z$ axes. Reproduced from Ref. \cite{McCVoorhees}, with the permission of AIP Publishing.}
\label{Fig1}
\end{figure}

After a growth process has ended, a section of a cylindrical monocrystalline wire of the radius $R_0$ is sitting on a substrate. A wire surface is making the angle $\alpha$ with the substrate 
at both contact lines, with $0<\alpha < \pi$. 
A set of cylindrical coordinates $(r,\theta,y)$ is introduced, where the angle $\theta$ is the angular coordinate measured 
counter-clockwise from the $x$-axis, and $y$ is the coordinate along the wire axis (Fig. \ref{Fig1}).  
The wire and the contact lines are infinite in the $y$ direction. A wire of constant radius $R_0$ is then described as $r=R_0$ for 
$\theta_0\le \theta \le \pi-\theta_0$, where $0\le\theta_0\le \pi/2$.  The contact lines for this wire are located at $\theta=\theta_0$ and at $\theta=\pi-\theta_0$.
Also, $\theta_0=\pi/2-\alpha$, if $0<\alpha\le\pi/2$, or $\theta_0=\alpha-\pi/2$, if $\pi/2<\alpha\le \pi$.
This wire is the \emph{equilibrium shape}.
A constant electric field $E_0$ is applied 
along the wire axis (the $y$-axis), inducing electromigration on the wire surface. Thus the wire surface evolves due to surface diffusion 
and electromigration.  At $\alpha=\pi$ a full-circular wire is free-standing, 
i.e. there is no contact 
with the substrate (in fact, it is near impossible to grow a full-circular wire on a substrate \cite{KTBECKN,KTEBCKN}). 
\begin{figure}[H]
\vspace{-0.2cm}
\centering
\includegraphics[width=4.0in]{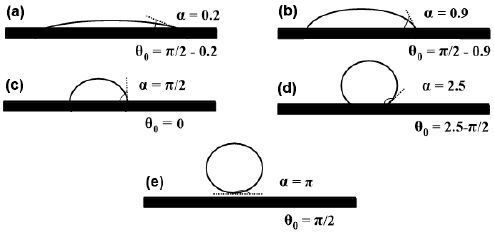}
\vspace{-0.15cm}
\caption{A sketch of equilibrium wire shapes at various contact angles.}
\label{Fig2}
\end{figure}

The goal of the following analysis is to determine the instability of the equilibrium shape, where the contact angle $\alpha$ is a prescribed constant.

Toward this goal, a wire is perturbed away from the equilibrium shape by an arbitrary perturbation. Let such perturbed wire be described by $r=F(t,\theta,y)$. 
Also let $\bm{e}_r,\ \bm{e}_\theta$, and $\bm{e}_y$ be the (pairwise orthogonal) unit vectors in the $r$, $\theta$, and $y$ directions, respectively, 
and let partial derivatives of $F$ be denoted by subscripts. With this notation, the position vector of the surface is written as 
$\bm{r}=(F\cos{\theta})\bm{e}_x+(F\sin{\theta})\bm{e}_z+y\bm{e}_y$, the normal velocity is $\bm{r}_t\cdot \bm{n}$ 
(where $\bm{n}$ is the unit outward normal to the surface), and the mean curvature of the surface is $\kappa=\bm{\nabla}\cdot \bm{n}$.
We also introduce the angle $\phi$ that quantifies the slope of the surface in $yz$ plane, the lengthscale $L$, and the timescale $\tau$:
$F_y=\tan{\phi}$, $L=R_0\sqrt{(\alpha-\sin{\alpha}\cos{\alpha})/\pi}$, $\tau=L^4 k T/(D\gamma\Omega^2\nu)$. Here, $k T$ is the thermal energy, 
$D$ the surface diffusivity of the adatoms, $\gamma$ the surface energy, $\Omega$ the atomic volume, and $\nu$ the surface density of the adatoms. 
The dimensionless variables are defined as follows: $\tilde r =r/L$, $\tilde y = y/L$, $\tilde \kappa = \kappa L$, $\tilde F = F/L$, $\tilde R =R_0/L$ and 
$\tilde \Delta = \Delta/L$. 

Dropping the tildas, the dimensionless equation describing the evolution of the wire surface reads:
\begin{equation}
\bm{r}_t\cdot \bm{n}=\bm{\nabla}_s\cdot\left[M(\phi)\left\{\bm{\nabla}_s\kappa+E\cos{(\phi)}\bm{e}_y\right\}\right],
\label{Vn}
\end{equation}
where
\begin{equation}
\bm{n}=\frac{F\bm{e}_r-F_\theta \bm{e}_\theta-F F_y\bm{e}_y}{q},\quad 
\kappa=\frac{1}{q}-\frac{\partial}{\partial y}\left(\frac{F F_y}{q}\right)+\frac{F_\theta^2}{q^3}-\frac{1}{F^2}\frac{\partial}{\partial \theta}\left(\frac{F F_\theta}{q}\right),
\quad \cos{\phi}=\frac{1}{\sqrt{1+F_y^2}},
\label{n_and_curv}
\end{equation}
with
\begin{equation}
q = \sqrt{F^2\left(1+F_y^2\right)+F_\theta^2}.
\label{q}
\end{equation}
Also
\begin{equation}
\bm{\nabla}_s = \frac{\left(1+F_y^2\right)\frac{\partial}{\partial \theta}-F_\theta F_y\frac{\partial}{\partial y}}{q^2}\bm{e}_\theta+\frac{\left(F^2+F_\theta^2\right)\frac{\partial}{\partial y}-F_\theta F_y \frac{\partial}{\partial \theta}}{q^2}\bm{e}_y
\label{nablas}
\end{equation}
is surface gradient (obtained in-house), and
\begin{equation}
M(\phi)=\frac{1+S\cos^2{\left[m\left(\phi+\psi\right)\right]}}{1+S},\quad \phi=\arctan{F_y}
\label{mobility}
\end{equation}
is the anisotropic diffusional mobility of the adatoms \cite{SK}. 
In Eq. (\ref{mobility}) $S$ is the anisotropy strength, $m=1,2,$ or $3$ the number of symmetry axes, and the misorientation angle $\psi$ 
is the angle between a crystalline symmetry direction, such as [110], [100], or [111], and the average surface orientation 
(i.e., the average orientation in the $yz$ plane of the unit normal to the surface). A crystalline symmetry direction 
also determines the number of symmetry axes.
For [110] direction: $m=1$, $0\le \psi\le \pi/2$; for [100] direction: $m=2$, $0\le \psi\le \pi/4$; 
for [111] direction: $m=3$, $0\le \psi\le \pi/6$ \cite{DM}. We call these angle intervals \emph{admissible}.
$E=Q E_0 L^2/(\Omega \gamma)=Q \Delta V L^2/(\Omega \gamma \ell)$,
where $Q>0$ is the effective charge of ionized atoms, $\Delta V$ applied voltage (to the front and back faces of the wire), and $\ell$ the wire length.
At the typical values $R_0=25$nm \cite{KTBECKN}, $\alpha=\pi/4$: $L=7.5$nm. Also, the typical $Q=10^{-9}$C \cite{REW}, 
$\gamma=2500$erg$/\mbox{cm}^2$, $\Omega=10^{-22}$cm$^3$, 
and $\ell=1000$nm \cite{KTBECKN}. Then at the representative value $\Delta V=1$V, $E=22.7$. 

Along the contact line at $\theta=\Theta(t,y)$ there are three dimensionless boundary conditions. 
\begin{enumerate}
\item The contact angle $\alpha$ is constant: 
\begin{equation}
\cos{\alpha}=\bm{n}\cdot \bm{e}_z;
\label{bc1}
\end{equation}
\item The contact lines lie in the plane of the substrate, or equivalently, the substrate surface is always $\Delta$ units above the origin of the 
cylindrical coordinate system, see Fig. \ref{Fig1}(b):  
\begin{equation}
F(t, \Theta, y)\sin{\Theta}=\Delta=R\sin{\theta_0};
\label{bc2}
\end{equation}
\item The total flux of adatoms on the wire surface along the direction of the tangent unit vector $\bm{m}$, see Fig. \ref{Fig1}(a) 
(into the substrate) is zero: 
\begin{equation}
\bm{m}\cdot \bm{\nabla}_s \left[M(\phi)\left(\kappa+E\cos{\phi}\right)\right]=\left[\bm{\nabla}_s M(\phi)\left(\kappa+E\cos{\phi}\right)\right]_\theta=0,
\label{bc3}
\end{equation}
where the subscript $\theta$ denotes the $\theta$-component of the surface gradient. 
\end{enumerate}  

\emph{Remark 3.}$\;$  Eq. (\ref{Vn}) is simply the expression for the normal velocity
of a wire surface; at $E=0$ and $S=0$, or $M(\phi)=1$ (isotropy of surface diffusion), this expression reduces to $\bm{r}_t\cdot \bm{n}=\bm{\nabla}_s^2 \kappa$, Eq. (1) in Ref. 
\cite{McCVoorhees}, i.e. the surface Laplacian of mean curvature, which is the hallmark of surface diffusion-driven morphological 
evolution \cite{Mullins95,CT94}. Also at $E=S=0$, the boundary condition (\ref{bc3}) reduces to 
$\bm{m}\cdot \bm{\nabla}_s \kappa=\left[\bm{\nabla}_s \kappa\right]_\theta=0$, which is Eq. (6) in Ref. \cite{McCVoorhees}. 
The surface diffusion chemical potential of adatoms is equal to the mean curvature $\kappa$, as follows from the analysis of the free-energy changes 
involved in the transfer of matter  along the surface \cite{Herring51}. It reminds the Laplace pressure at a fluid surface. 

\emph{Remark 4.}$\;$ The dimensionless $R=\sqrt{\pi/(\alpha-\sin{\alpha}\cos{\alpha})}$, which is the monotonically decreasing function of $\alpha$.

\section{Linear stability of the equilibrium wire shape}
\label{LSA}

Consider small perturbations of the shape and contact-line positions as follows:
\begin{equation}
r=F=R+\epsilon \hat r(t,\theta,y), \Theta=\theta_0+\epsilon \hat \theta(t,y),
\label{perturb}
\end{equation}
where $0<\epsilon\ll 1$. Substitute Eqs. (\ref{perturb}) in the system (\ref{Vn})-(\ref{bc3}) and keep only terms of order $\epsilon$. 
Eqs. (\ref{Vn})-(\ref{mobility}) yield the linear equation for $\hat r$:
\begin{equation}
R^4(1+S) \hat r_t = -\frac{2+S(1+\cos{2m\psi})}{2}\left[R^4 \hat r_{yyyy}+R^2\left(\hat r_{yy}+2\hat r_{\theta \theta yy}\right)+\hat r_{\theta \theta}+
\hat r_{\theta \theta \theta \theta}\right]-E m S R^4 \sin{(2m\psi)}\hat r_{yy}.
\label{eq_hat_r}
\end{equation}
With the help of the linearized boundary condition (\ref{bc2}), $\tan{(\theta_0)}\hat r+\hat \theta=0$, the boundary conditions (\ref{bc1}) and (\ref{bc3}) can be reduced \cite{McCVoorhees} to (i) a linearized contact angle conditions 
\begin{equation}
\mbox{At}\;\; \theta=\theta_0:\quad \hat r_\theta + \tan{\left(\theta_0\right)}\hat r=0;\qquad \mbox{At}\;\; \theta=\pi- \theta_0:\quad \hat r_\theta - \tan{\left(\theta_0\right)}\hat r=0
\label{bc_set1}
\end{equation} 
and (ii) a linearized no-flux conditions:
\begin{equation}
\mbox{At}\;\; \theta=\theta_0,\; \theta=\pi- \theta_0:\quad 
[2+S(1+\cos{2m\psi})]\left(R^2 \hat r_{\theta yy}+\hat r_\theta +\hat r_{\theta \theta \theta}\right)
+2R^2 m S \sin{(2m\psi)}\left(\frac{1}{R}+E\right)\hat r_{\theta y}=0. \label{bc_set2}
\end{equation}
At $S=0$ Eqs. (\ref{eq_hat_r}) and (\ref{bc_set2}) can be seen to yield Eqs. (9) and (11) in  Ref. \cite{McCVoorhees}. 
Also, Eqs. (\ref{eq_hat_r}) and (\ref{bc_set2}) show that at the misorientation angles $\psi=n\pi/2m$ the effect of the applied electric field 
vanishes, but the effect of the anisotropy persists.
Indeed, for instance, at $\psi=0$ a crystalline symmetry direction, i.e. the direction of fast surface diffusion, is collinear to the average 
orientation in the $yz$ plane of the unit normal to the surface, and therefore the direction of fast surface diffusion is orthogonal to the projection of the 
applied electric field onto the surface, $E\cos{(\phi)}\bm{e}_y$.

Since the perturbation $\hat \theta$ does not enter the linearized problem (\ref{eq_hat_r})-(\ref{bc_set2}),  a solution of that problem is sought in the form
\begin{equation}
\hat r(t,\theta,y)=\bar r(\theta)\exp{(\sigma t + iky)},
\label{hatr}
\end{equation}
where $\bar r(\theta)$ is an unknown function to be determined, $k$ the axial wavenumber, and $\sigma$ the growth rate. 

For axisymmetric perturbations $(\bar r(\theta)=const.)$ the problem (\ref{eq_hat_r})-(\ref{bc_set2}) is well-posed only for $\alpha =\pi$ and $\pi/2$.
Indeed, at $\alpha\neq \pi, \pi/2$ the boundary conditions (\ref{bc_set2}) are identically zero after application of Eq. (\ref{hatr}) with 
$\bar r(\theta)=const.$, but the boundary conditions (\ref{bc_set1}) read $\hat r=0$. 
The latter conditions can't be applied, since Eq. (\ref{eq_hat_r}) after applying Eq. (\ref{hatr}) with $\bar r(\theta)=const.$ immediately yields the growth rate. 
At $\alpha=\pi$ a wire is free-standing and the boundary conditions (\ref{bc_set1}), (\ref{bc_set2}) are not needed.
At $\alpha=\pi/2$ a wire is semi-cylindrical, i.e. it is the upper-half 
of a cylindrical wire that makes a $90^\circ$ contact angle with the substrate (Fig. \ref{Fig2}). In this case the boundary conditions 
(\ref{bc_set1}), (\ref{bc_set2}) are identically zero.
To summarize, for axisymmetric perturbations at $\alpha=\pi, \pi/2$ the growth rate $\sigma$, the instability cutoff wavenumber $k_c$, and the most dangerous wavenumber $k_m$ 
(i.e., the one that maximizes the growth rate) follow from Eq. (\ref{eq_hat_r}):
\begin{equation}
\sigma=\frac{2+S(1+\cos{2m \psi})}{2R^2(1+S)}k^2\left(1-R^2k^2\right)+E \frac{m S \sin{2m \psi}}{1+S}k^2,
\label{sigma_axi}
\end{equation} 
\begin{equation}
k_c=\frac{1}{R}\sqrt{1+E\frac{2m R^2 S \sin{2m \psi}}{2+S(1+\cos{2m \psi})}},\quad k_m=\frac{k_c}{\sqrt{2}},
\label{kc_axi}
\end{equation}
where $R=1,\ \sqrt{2}$ at $\alpha=\pi,\ \pi/2$, respectively. 
Artificially assuming that Eq. (\ref{kc_axi}) holds for all contact angles of interest, i.e. $0<\alpha\le \pi$, we see that $k_c(E=0)=1/R=\sqrt{(\alpha-\sin{\alpha}\cos{\alpha})/\pi}$, 
thus without the electric field $k_c$ for axisymmetric perturbations would monotonically increase with the contact angle $\alpha$ (see Fig. \ref{kc_atE=0_m=13}).\footnote{For wires that make a contact with the substrate, McCallum \emph{et al.} \cite{McCVoorhees} make the following useful observation: ``...at any other contact
angle (i.e. at $\alpha\ne \pi/2$), an axisymmetric perturbation does not preserve the
contact line; it will lead either to elevation of the contact line
out of the plane of the substrate, or to penetration of the
contact line below the plane of the substrate. In other words,
the outward normal $\bm{n}$ lies in the plane of the substrate only
when $\alpha=\pi/2$, so that an axisymmetric perturbation at this
contact angle can keep the contact line in the substrate plane".}
The axisymmetric case of a free-standing wire deserves special attention, and we discuss it in Sec. \ref{Res1}.

For non-axisymmetric perturbations, the
substitution of Eq. (\ref{hatr}) into Eqs. (\ref{eq_hat_r})-(\ref{bc_set2}) results in a system of linear algebraic equations $A\bm{c}=0$, 
where $\bm{c}$ is a four-component real vector, and the four-by-four matrix $A$ is complex at $S\neq 0$ (see Appendix). This complexity stems from the terms proportional to $\hat r_{\theta y}$ in Eq. (\ref{bc_set2}).
We used Mathematica\,\textsuperscript{\tiny\textregistered} to find $A$. At $S=0$ that matrix simplifies to a real matrix that can be seen in Ref. \cite{McCVoorhees}. 
For nontrivial solutions we need $\mbox{Re}(\mbox{Det}(A))=0$.
This equation is the characteristic equation for our problem, because it is a functional constraint that must be satisfied by the seven
parameters $\sigma$,  $k$, $\alpha$, $E$, $S$, $m$, and $\psi$. Due to complexity of that equation we are not presenting it in this paper. Moreover, 
that equation defies the analytical solution for 
$\sigma$ in terms of $k$ and other parameters due to excess time involved in algebraic transformations in Mathematica\,\textsuperscript{\tiny\textregistered}. However, the neutral stability can be determined by 
setting $\sigma=0$. The resultant equation of neutral stability $\mbox{Re}[\mbox{Det}(A(\sigma=0,k,\alpha,E,S, m,\psi))]=0$ has the same general form as Eq. (20) in 
Ref. \cite{McCVoorhees}:
\begin{equation}
a_2\tan^2{\alpha}+a_1\tan{\alpha}+a_0=0,
\label{Eq20McCVoorhees}
\end{equation}
where $a_0,\ a_1$ and $a_2$ are the functions of $k,\alpha,E,S, m,\psi$. These functions are very cumbersome, thus they are not presented in this paper 
(a Mathematica\,\textsuperscript{\tiny\textregistered} notebook 
with these functions is available on request from the author).
At $S=0$ Eq. (\ref{Eq20McCVoorhees}) reduces to Eq. (20) in Ref. \cite{McCVoorhees}, and in this case the neutral stability is analyzed in that paper in detail.
Moreover, the functions $a_0, a_1$ and $a_2$ are non-divergent at $\alpha=0$ and $\alpha=\pi/2$ and therefore at these values of $\alpha$ Eq. (\ref{Eq20McCVoorhees}) reduces to 
equations $a_0(k,\alpha=0,E,S, m,\psi)=0$ and $a_2(k,\alpha=\pi/2,E,S, m,\psi)=0$, respectively. These equations can be analytically solved for 
$k(E, S, m,\psi)\equiv k_c(E, S, m,\psi)$ (where $k_c$ denotes the cutoff wavenumber of a longwave instability), 
and therefore at $\alpha=0$ and $\alpha=\pi/2$ the neutral stability curve can be explicitly found in the, say, $E-k_c$ plane. The case $\alpha=0$ is of no interest from a practical point of view (since there is no wire), but 
the (non-axisymmetric) case $\alpha=\pi/2$ will be briefly discussed below.
At $\alpha \ne \pi/2$ we find the neutral stability curves by fixing all parameters except $k$ and one 
of $\alpha$, $E$, $S$, $m$, and $\psi$, and then computing the zero level set of a two-variable function $a_2\tan^2{\alpha}+a_1\tan{\alpha}+a_0$. 

\section{Results}
\label{Res}

First, it is obvious that larger anisotropy, i.e. larger $S$, results in a decrease of stability. Thus  
we focus on effects of variations of $\alpha$ and $m$ at fixed $S$ and $\psi$. Thus for the rest of the paper we fix $S=1$, since the typical values of $S$ in the literature are of 
order 1, and $\psi=\pi/12$, since this small non-zero angle is common to all three admissible intervals \cite{DM,SK}.

\subsection{Special case I: a free-standing wire ($\alpha=\pi$), axisymmetric perturbations}
\label{Res1}

\begin{figure}[h]
\vspace{-0.2cm}
\centering
\includegraphics[width=3.0in]{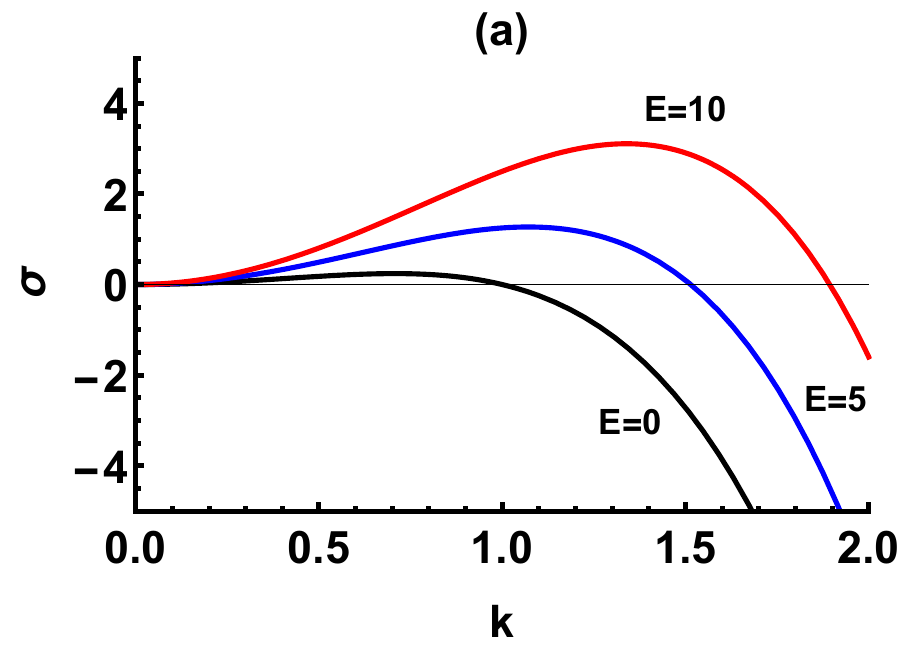}\hspace{0.5cm}
\includegraphics[width=3.0in]{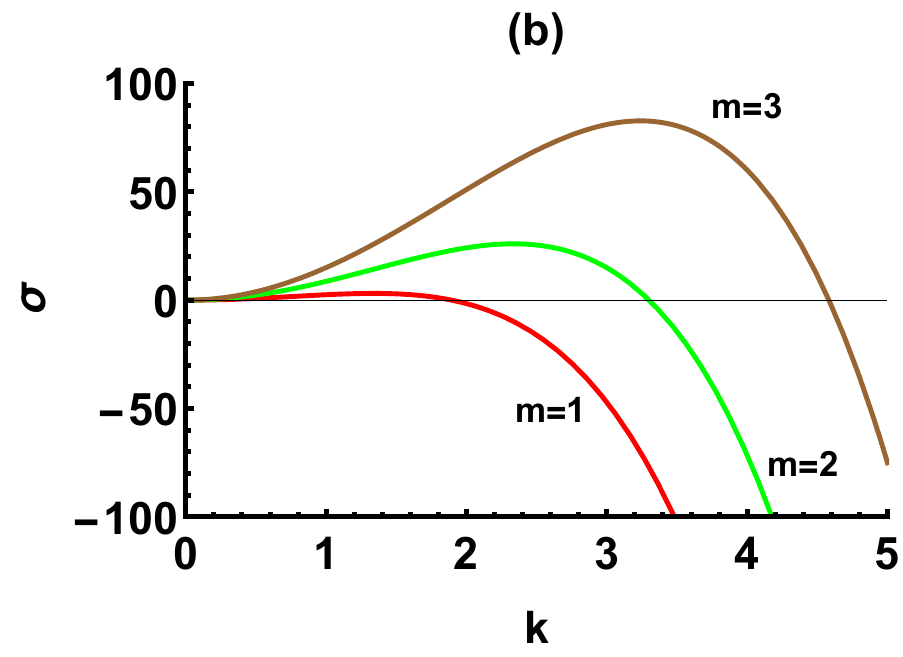}\hspace{0.5cm}\\
\vspace{-0.15cm}
\caption{Growth rate $\sigma$ for a free-standing wire, $\alpha=\pi$. See Eq. (\ref{sigma_axi_alpha_pi}).
(a) $m=1$. 
(b) $E=10$. The red curve is the same in both panels.   
}
\label{sigma}
\end{figure}

In this case $R=1$, therefore Eqs. (\ref{sigma_axi}) and (\ref{kc_axi}) give:
\begin{equation}
\sigma=\frac{2+S(1+\cos{2m \psi})}{2(1+S)}k^2\left(1-k^2\right)+E \frac{m S \sin{2m \psi}}{1+S}k^2,
\label{sigma_axi_alpha_pi}
\end{equation}
\begin{equation}
k_c=\sqrt{1+E\frac{2m S \sin{2m \psi}}{2+S(1+\cos{2m \psi})}},\quad k_m=\frac{k_c}{\sqrt{2}}.
\label{kc_axi_alpha_pi}
\end{equation}
Without anisotropy, i.e. at $S=0$ Eq. (\ref{sigma_axi_alpha_pi}) yields $\sigma=k^2\left(1-k^2\right)$, which is the growth rate obtained by 
Nichols and Mullins \cite{NM2}.
Anisotropy alone, without the electric field, results in the multiplication of this $\sigma$ by the factor $u(S,m,\psi)\equiv[2+S(1+\cos{2m \psi})]/[2(1+S)]$, 
that does not depend on $k$.
Thus with or without anisotropy, at $E=0$ the non-trivial solution of $\sigma=0$ is the cutoff wavenumber $k_c=1$ and the most dangerous wavenumber, 
i.e. the one that maximizes the growth rate, 
is $k_m=k_c/\sqrt{2}=1/\sqrt{2}$. These values are the same as for the classical longwave Rayleigh-Plateau instability. The anisotropy at $E=0$ only slightly 
affects the maximum growth rate $\sigma_{m}$; for instance, 
$u(1,1,\pi/12)=0.9665$. 
In the presence of the electric field, 
since $0\le 2m \psi\le \pi$, the coupling of the electric field and anisotropy results in instability enhancement, i.e. in larger $k_c$, $k_m$, 
and $\sigma_{m}$, as seen in Fig. \ref{sigma}(a). Also, at fixed $E$, larger $m$ has the same effect (Fig. \ref{sigma}(b)). In dimensional coordinates, 
$k_c>1$ at $E>0$ implies that a wire is destabilized by perturbations that have wavelengths shorter than a wire circumference $2\pi R_0$.

\subsection{Special case II: $\alpha=\pi/2$, non-axisymmetric perturbations}
\label{Res2}

Remark: In this section and in the next section, note that in all Figures an instability region is below a neutral stability curve; the stability region is
therefore above a neutral stability curve.

As pointed out in section \ref{LSA}, in this case even for non-axisymmetric 
perturbations the neutral stability curves are the analytical explicit solutions of equation $a_2(k,\alpha=\pi/2,E,S=1, m,\psi=1/12)=0$. Also $R=\sqrt{2}$.

\begin{figure}[H]
\vspace{-0.2cm}
\centering
\includegraphics[width=3.0in]{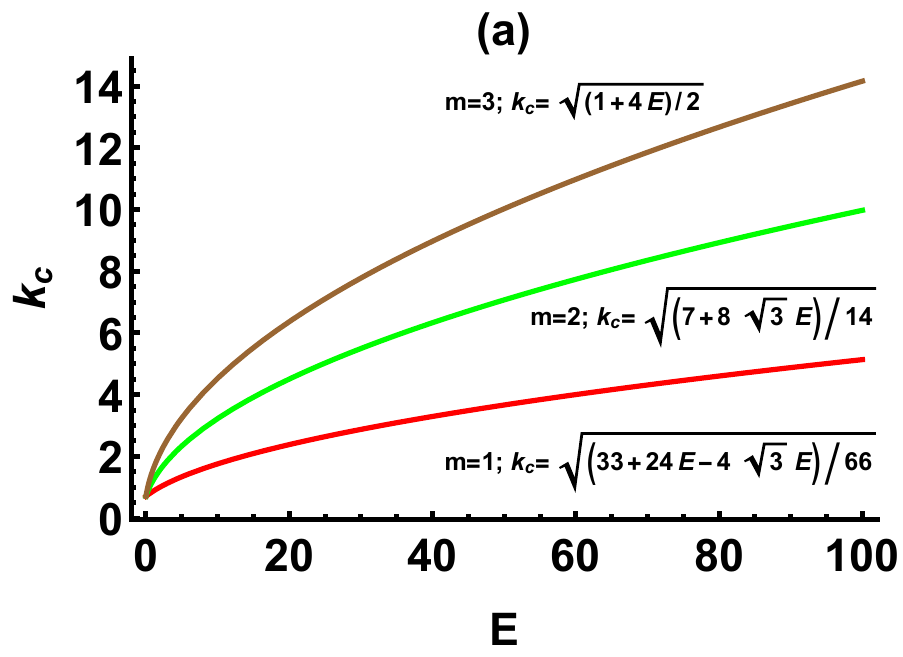}\hspace{0.5cm}
\includegraphics[width=3.0in]{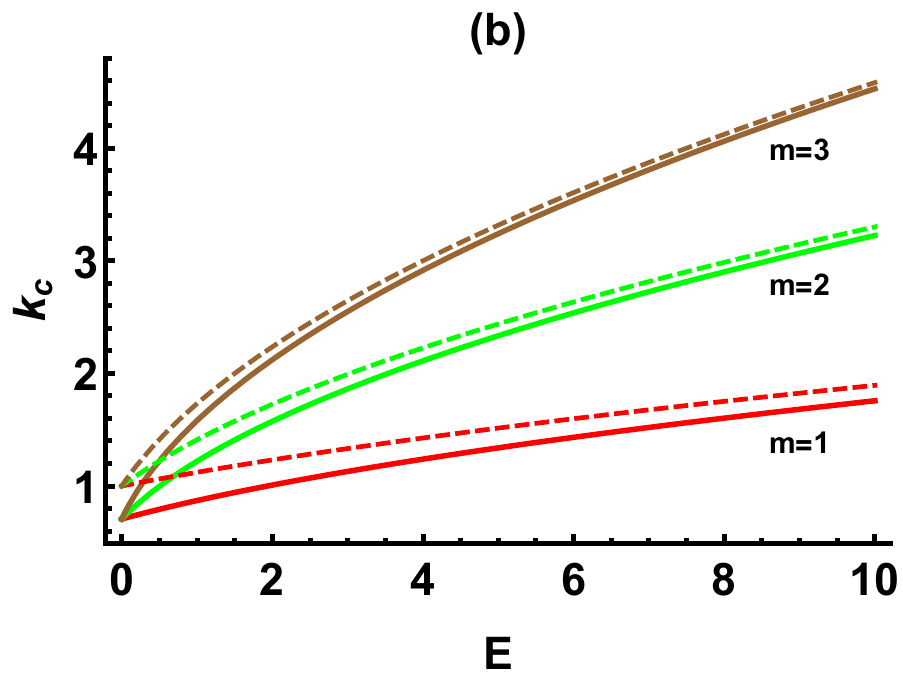}\hspace{0.5cm}\\
\vspace{-0.15cm}
\caption{(a) The neutral stability curves at $\alpha=\pi/2$. Here, $k_c(0)=1/R=1/\sqrt{2}$. (b) Comparison of neutral stability for $\alpha=\pi/2$ to neutral stability of a
free-standing wire. The latter curves are dashed.} 
\label{Fig_m=123}
\end{figure}

Fig. \ref{Fig_m=123}(a) shows these analytically determined neutral stability curves for $m=1,2,3$.  
Larger $m$ results in a larger instability region. 
Interestingly, it is easy to see from Eq. (\ref{kc_axi}) that equations of 
axisymmetric neutral stability curves coincide with equations that are marked in Fig. \ref{Fig_m=123}(a), i.e. the neutral stability at $\alpha=\pi/2$ is 
insensitive to the symmetry (or to the lack of symmetry) of perturbations. 

In Fig. \ref{Fig_m=123}(b) the comparison of two axisymmetric cases,  $\alpha=\pi/2, \pi$ is shown. 
At each of the three $m$ values, the larger values of a cut-off wavenumber  
for a free-standing wire in a weak electric field signal that a full-circular free-standing wire is more unstable than a semi-circular 
one of the same radius on a substrate. Note also that for either wire, the distance between
$m=1, 2, 3$ neutral stability curves increases as $E$ increases. This means that a wire crystallographic orientation on 
a substrate has larger impact on stability in stronger applied electric fields.    

%
%

%

\subsection{General case: $0< \alpha< \pi$, non-axisymmetric perturbations}
\label{Res3}

In this section we determine the neutral stability curves by computing the zero level set of a two-variable function 
$a_2\tan^2{\alpha}+a_1\tan{\alpha}+a_0$ (Eq. (\ref{Eq20McCVoorhees})). 

\begin{figure}[h]
\vspace{-0.2cm}
\centering
\includegraphics[width=3.0in]{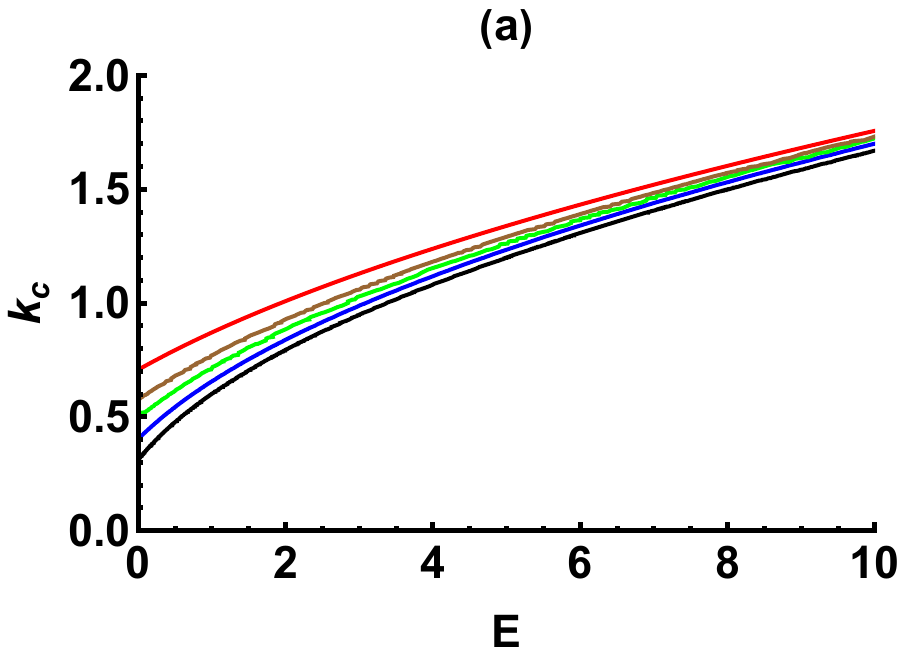}\hspace{0.5cm}
\includegraphics[width=3.0in]{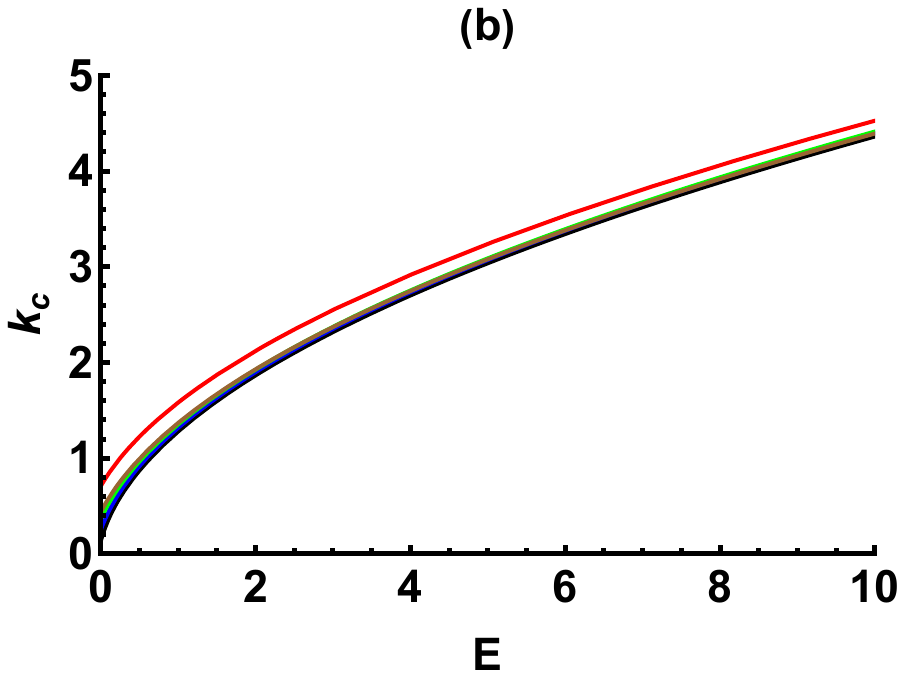}\hspace{0.5cm}\\
\vspace{-0.15cm}
\caption{(a) Neutral stability curves at $m=1$, from bottom to top: $\alpha=18^\circ,\ 30^\circ,\ 45^\circ,\ 60^\circ,\ 90^\circ$. The top red curve ($\alpha=90^\circ$) 
is the copy of the red curve from Fig. \ref{Fig_m=123}(a); this is the analytical result. 
(b) Same as (a), but for $m=3$. The top red curve ($\alpha=90^\circ$) 
is the copy of the brown curve from Fig. \ref{Fig_m=123}(a); this is the analytical result.} 
\label{Fig_AcuteAlpha_m=13}
\end{figure}

Figures \ref{Fig_AcuteAlpha_m=13}(a,b) show the neutral stability curves for acute contact angles, $0^\circ< \alpha< 90^\circ$ and $m=1,\ 3$. As expected, the size of the instability 
region increases as $\alpha$ increases, albeit the dependence is weak for $m=1$ and very weak for $m=3$. Notice that all instability regions at $m=1$ are smaller than any 
instability region at $m=3$, thus at $m=3$ the wire is overall more unstable than at $m=1$. But at $E\lessapprox 0.5$, $k_c(m=1)>k_c(m=3)$, 
thus at a weak electric field a wire is more stable at $m=3$.

\begin{figure}[h]
\vspace{-0.2cm}
\centering
\includegraphics[width=3.0in]{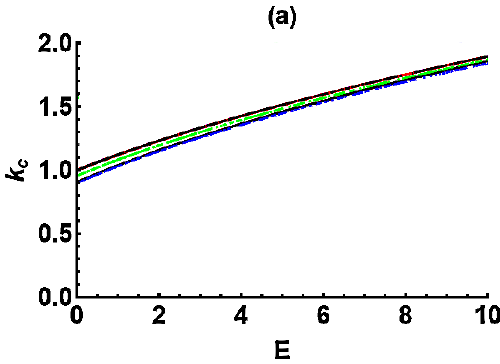}\hspace{0.5cm}
\includegraphics[width=2.0in]{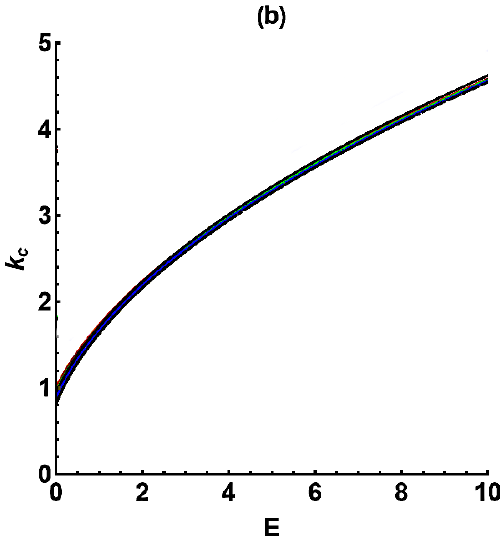}\hspace{0.5cm}\\
\vspace{-0.15cm}
\caption{(a) Neutral stability curves at $m=1$, from bottom to top: $\alpha=105^\circ,\ 120^\circ,\ 135^\circ,\ 160^\circ,\ 170^\circ,\ 175^\circ$.
(b) Same as (a), but at $m=3$. Broken curves in the panel (a) is the artifact of a numerical solution. } 
\label{Fig_ObtuseAlpha_m=13}
\end{figure}

Figures \ref{Fig_ObtuseAlpha_m=13}(a,b) show the neutral stability curves for obtuse contact angles, $90^\circ< \alpha< 180^\circ$ and $m=1,\ 3$. Here, the size of the instability 
region shows very weak dependence on the contact angle for both values of $m$. 
Due to large numerical errors we were unable to compute the curves for very large values of $\alpha$, $175^\circ < \alpha < 180^\circ$. 
The curves computed at $\alpha=175^\circ$ essentially coincide with the analytical solutions for a free-standing wire ($\alpha=\pi$) and axisymmetric perturbations, Eq. (\ref{kc_axi}).

In Figures \ref{Fig_AcuteAlpha_m=13} and \ref{Fig_ObtuseAlpha_m=13} we also observe that in a strong applied electric field the difference in stability
for various contact angles vanishes. This is similar to Fig. \ref{Fig_m=123}(b), that shows the comparison between the axisymmetric cases $\alpha=\pi/2, \pi$.
Thus the destabilizing effect of a larger contact angle vanishes in comparison to the destabilizing effect of the electic field at $E\sim 20$.

\begin{figure}[h]
\vspace{-0.2cm}
\centering
\includegraphics[width=3.5in]{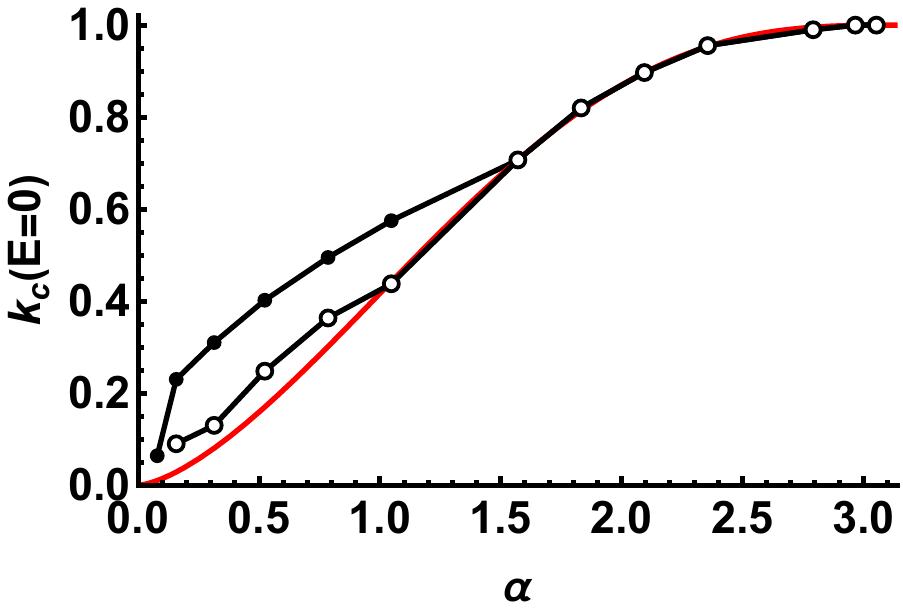}\hspace{0.5cm}
\vspace{-0.15cm}
\caption{$k_c$ at $E=0$. Solid circles: $m=1$, open circles: $m=3$. Black curves are only the guide for the eye. 
Red curve: Eq. (\ref{kc_axi}) at $E=0$.} 
\label{kc_atE=0_m=13}
\end{figure}

Last, Fig. \ref{kc_atE=0_m=13} shows $k_c(E=0)$ as the function of the contact angle. The data was extracted from Figures \ref{Fig_AcuteAlpha_m=13}
and \ref{Fig_ObtuseAlpha_m=13}. The comparison in Fig. \ref{kc_atE=0_m=13} is to axisymmetric perturbations, i.e. Eq. (\ref{kc_axi}) at $E=0$ 
(assuming, artificially, that this equation holds for all angles $0^\circ <\alpha\le 180^\circ$).
For either $m$, $k_c(E=0)$ for a non-axisymmetric perturbation is larger than the one for an axisymmetric perturbation up to a threshold $\alpha$ value,
which is $60^\circ$ at $m=3$ and $90^\circ$ at $m=1$. On the scale of the Figure, at the contact angles larger than $60^\circ$ ($90^\circ$) there is 
no difference in $k_c(E=0)$ between axisymmetric and non-axisymmetric perturbations, and between non-axisymmetric perturbations at $m=1$ and at $m=3$. 
And again it is seen here that larger contact angles yield weaker stability (since $k_c$ increases as $\alpha$ increases).  

\section{Summary and discussion}
\label{Conc}

The linear stability analysis in this paper shows that wires that are oriented along [110] crystalline direction are most stable, 
and those oriented along [111] direction are most unstable. In applied axial electric field a wire is destabilized by perturbations that have wavelengths 
shorter than a wire circumference.
Stronger electric field results in a wider interval of unstable wavelengths, a decrease of a most dangerous unstable wavelength, and an increase of a 
maximum growth rate of the instability.
For a typical nanowire the destabilizing effect of a larger contact angle 
vanishes in comparison to the destabilizing effect of the electromigration at the electrical potential difference of the order of one to ten volts.
Also, a wire crystallographic orientation on 
a substrate has larger impact on stability in a strong electric field.
Non-axisymmetric perturbations destabilize a wire on a substrate, which contrasts a free-standing wire;
for the latter, non-axisymmetric perturbations decay \cite{NM2}.

\begin{figure}[h]
\vspace{-0.2cm}
\centering
\includegraphics[width=3.5in]{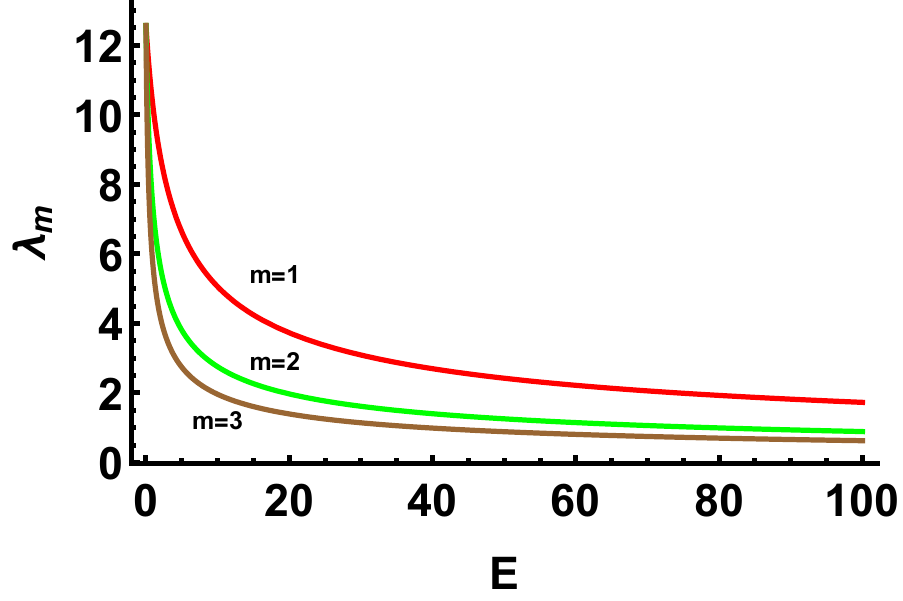}\\
\vspace{-0.15cm}
\caption{Most dangerous wavelength for the $90^\circ$ contact angle (Eq. (\ref{lambda_m})).
}
\label{lmbda_m}
\end{figure}

The one parameter of this morphological stability study that is of keen interest to applications is $\lambda_m$, 
i.e. the most dangerous unstable wavelength. This parameter was extensively correlated to the distance between 
the centers of a spherical nanoparticles that emerge after a wire breakup (a cylindrical sections that emerge immediately after a breakup 
quickly evolve into spheroids under the action of a volume-minimizing surface diffusion flow) \cite{KTBECKN,KTEBCKN,LBPDVC,KT}. 
To better emphasize that $\lambda_m$ decreases as a result of electromigration, in Fig. \ref{lmbda_m} it is is plotted vs. $E$
at $\alpha=\pi/2$. At this contact angle and regardless of the perturbations' symmetry, $\lambda_m$ is given by (see Eq. (\ref{kc_axi}), where $R=\sqrt{2}$):
\begin{equation}
\lambda_m=\frac{2\pi \sqrt{2}}{k_c}=\frac{4\pi}{\sqrt{1+E\frac{4m S \sin{2m \psi}}{2+S(1+\cos{2m \psi})}}}.
\label{lambda_m}
\end{equation}
Moreover, it is seen that applying the electric field to a wire that has been grown along a special crystallographic orientation allows to fine tune 
the spacing between the nanoparticles and therefore
exercise control over aerial density of nanoparticles on the substrate.

And finally, we point out that surface electromigration may stabilize a wire that undergoes a Rayleigh-Plateau instability, thus suppressing a wire breakup.
Indeed, switching the polarity of the electrical contacts is equivalent to changing the sign of the applied voltage difference, i.e. $\Delta V \rightarrow -\Delta V$. 
This gives $E<0$. Then, at $\alpha=\pi/2$ it follows from Eq. (\ref{kc_axi}) with $R=\sqrt{2}$, that $\sigma < 0$ for all wavenumbers, if
\begin{equation}
E < E_{\alpha=\pi/2}=\frac{-\left[2+S(1+\cos{2m \psi})\right]}{4 m S \sin{2m \psi}}.
\label{Ecrit_pid2}
\end{equation}
And for free-standing wire Eq. (\ref{sigma_axi_alpha_pi}) and Fig. \ref{Fig_ObtuseAlpha_m=13} (see the discussion of that figure) 
give stability for all wavenumbers, if
$E < E_{\alpha=\pi}=2 E_{\alpha=\pi/2}$. For other contact angles the critical value of $E$ may be obtained numerically if required.  
Similar suppression of morphological instability by application of electric current was proposed for planar films evolving by surface diffusion 
(at isotropic surface energy, as in this paper) \cite{SGM,SGM1,DM}. 
Without electromigration, suppression of morphological instability was proposed for isotropic wires evolving by surface diffusion on patterned 
substrates \cite{BB}.

\setcounter{equation}{0}
\section{Appendix}
\label{App}

In Sec. \ref{LSA}, the matrix of a system of linear algebraic equations is: 
\begin{equation}
A=
\begin{pmatrix}
p_1 e^{p_1 \theta_0}b_1 & -p_1 e^{-p_1 \theta_0}b_1 & p_2 e^{p_2 \theta_0}b_2 & -p_2 e^{-p_2 \theta_0}b_2 \\
p_1 e^{p_1 \left(\pi-\theta_0\right)}b_1 & -p_1 e^{-p_1 \left(\pi-\theta_0\right)}b_1 & p_2 e^{p_2 \left(\pi-\theta_0\right)}b_2 & -p_2 e^{-p_2 \left(\pi-\theta_0\right)}b_2 \\
e^{p_1 \theta_0}\left(p_1+\tan{\theta_0}\right) & -e^{-p_1 \theta_0}\left(p_1-\tan{\theta_0}\right) & e^{p_2 \theta_0}\left(p_2+\tan{\theta_0}\right) & -e^{-p_2 \theta_0}\left(p_2-\tan{\theta_0}\right) \\
e^{p_1 \left(\pi-\theta_0\right)}\left(p_1-\tan{\theta_0}\right) & -e^{-p_1 \left(\pi-\theta_0\right)}\left(p_1+\tan{\theta_0}\right) & e^{p_2 \left(\pi-\theta_0\right)}\left(p_2-\tan{\theta_0}\right) & -e^{-p_2 \left(\pi-\theta_0\right)}\left(p_2+\tan{\theta_0}\right) 
\end{pmatrix}
, 
\label{matrix}
\end{equation}
where 
\begin{equation}
b_1=k_1+i k_2+k_3p_1^2,\ b_2=k_1+i k_2+k_3p_2^2, 
\label{bs}
\end{equation}
\begin{equation}
k_1=\frac{(2+S(1+\cos{2m\psi}))\left(k^2R^2-1\right)}{2R^4(1+S)},\ k_2=\frac{-k m S(1+R E)\sin{2m\psi}}{R^3(1+S)},\ k_3=\frac{-(2+S(1+\cos{2m\psi}))}{2R^4(1+S)},
\label{ks}
\end{equation}
\begin{equation}
p_1=\sqrt{\eta-\xi},\ p_2=\sqrt{\eta+\xi},\ \eta=-\frac{1}{2}+k^2 R^2,\ \xi=\frac{1}{2}\sqrt{1-\frac{8R^4\left(\sigma(1+S)-E m S k^2 \sin{2m\psi}\right)}{2+S(1+\cos{2m\psi})}}.
\label{ps}
\end{equation}

\vspace{0.5cm}
\noindent
{\bf ACKNOWLEDGMENTS}
\vspace{0.5cm}

\noindent
Prof. Tilak Bhattacharya is acknowledged for his help in deriving Eq. (\ref{nablas}).


\begin{thebibliography}{200}

\bibitem{NM2} F.A. Nichols and W.W. Mullins, ``Surface-(Interface-) and volume-diffusion contributions to morphological changes driven by capillarity", 
\textit{Trans. Metall. Soc. AIME}$\;$ {\bf 233},  1840–1848 (1965).

\bibitem{NM1} F.A. Nichols and W.W. Mullins, ``Morphological Changes of a Surface of Revolution due to Capillarity Induced Surface Diffusion",
\textit{J. Appl. Phys.}$\;$ {\bf 36}, 1826 (1965).

\bibitem{C} J.W. Cahn, ``Stability of rods with anisotropic surface free energy", 
\textit{Scripta Metall.}$\;$ {\bf 13},  1069–1071 (1979).

\bibitem{McCVoorhees} M.S. McCallum, P.W. Voorhees, M.J. Miksis, S.H. Davis, and H. Wong, ``Capillary instabilities in solid thin films: Lines", 
\textit{J. Appl. Phys.}$\;$ {\bf 79}, 7604 (1996).

\bibitem{BBW} A.J. Bernoff, A.L. Bertozzi, and T.P. Witelski, ``Axisymmetric surface diffusion: dynamics and stability of self-similar pinchoff", 
\textit{J. Stat. Phys.}$\;$ {\bf 93} 725–776 (1998).

\bibitem{WMVD} H. Wong, M.J. Miksis, P.W. Voorhees, and S.H. Davis, ``Universal pinch off of rods by capillarity-driven surface diffusion", 
\textit{Scripta Mater.}$\;$ {\bf 39}, 55–60 (1998).

\bibitem{KDMV} D.J. Kirill, S.H. Davis, M.J. Miksis, and P.W. Voorhees, ``Morphological instability of a whisker", 
\textit{Proc. R. Soc. Lond. A}$\;$ {\bf 455}, 3825-3844 (1999).

\bibitem{GM} K.F. Gurski, and G.B. McFadden, ``The effect of anisotropic surface energy on the Rayleigh instability", 
\textit{Proc. R. Soc. Lond. A}$\;$ {\bf 459} 2575–2598 (2003).

\bibitem{GMM} K.F. Gurski, G.B. McFadden, and M.J. Miksis, ``The effect of contact lines on the Rayleigh instability with anisotropic surface energy", 
\textit{SIAM J. Appl. Math.}$\;$ {\bf 66}, 1163–1187 (2006).

\bibitem{MyJCP} P.Du, M. Khenner, and H. Wong, ``A tangent-plane marker-particle method for the computation of three-dimensional solid surfaces evolving 
by surface diffusion on a substrate",
\textit{J. Comp. Phys.}$\;$ {\bf 229}, 813-827 (2010).





\bibitem{KTBECKN} S. Karim, M.E. Toimil-Molares, A.G. Balogh, W. Ensinger, T.W. Cornelius, E.U. Khan, and R. Neumann, 
``Morphological evolution of Au nanowires controlled by Rayleigh instability", \textit{Nanotechnology}$\;$ {\bf 17},  5954–5959 (2006).

\bibitem{KTEBCKN} S. Karim, M.E. Toimil-Molares,  W. Ensinger, A.G. Balogh, T.W. Cornelius, E.U. Khan, and R. Neumann, 
``Influence of crystallinity on the Rayleigh instability of gold nanowires", \textit{J. Phys. D.: Appl. Phys.}$\;$ {\bf 40}, 3767–3770 (2007).


\bibitem{LBPDVC} H. Lia, J.M. Biser, J.T. Perkins, S. Dutta, R.P. Vinci, and H.M. Chan, 
``Thermal stability of Cu nanowires on a sapphire substrate", \textit{J. Appl. Phys.}$\;$ {\bf 103}, 024315 (2008).

\bibitem{KT} G.H. Kim and C.V. Thompson, 
``Effect of surface energy anisotropy on Rayleigh-like solid-state dewetting and nanowire stability", \textit{Acta Mater.}$\;$ {\bf 84}, 190-201 (2015).



\bibitem{XLL} S. Xu, P.F. Li, and Y. Lu, ``In situ atomic-scale analysis of Rayleigh instability in ultrathin gold nanowires",
\textit{Nano Res.}$\;$ {\bf 11}, 625 (2018).








\bibitem{BTS} W. Bao, C.V. Thompson, and D.J. Srolovitz, ``Phase field approach for simulating solid-state dewetting problems",
\textit{Acta Materialia}$\;$ {\bf 60}, 5578-5592 (2012).


\bibitem{WZSB} Y. Wang, Q. Zhao, D.J. Srolovitz, and W. Bao, ``Solid-state dewetting and island morphologies in strongly anisotropic materials",
\textit{Scripta Materialia}$\;$ {\bf 115}, 123-127 (2016).

\bibitem{JWZ} W. Jiang, Y. Wang, and Q. Zhao, ``A parametric finite element method for solid-state dewetting problems with anisotropic surface energies",
\textit{J. Comp. Phys.}$\;$ {\bf 330}, 380-400 (2017). 

\bibitem{BRTMP} F. Boccardo, F. Rovaris, A. Tripathi, F. Montalenti, and O. Pierre-Louis,  ``Stress-Induced Acceleration and Ordering in Solid-State Dewetting",
\textit{Phys. Rev. Lett.}$\;$ {\bf 128}, 026101 (2022).


\bibitem{EWCWGBCCT} M.A. L'Etoile, B.M. Wang, Q. Cumston, A.P. Warren, J.C. Ginn, K. Barmak, K.R. Coffey, W.C. Carter, and C.V. Thompson, 
``Experimental and Computational Study of the Orientation Dependence of Single-Crystal Ruthenium Nanowire Stability",
\textit{Nano Lett.}$\;$ {\bf 22}, 9958 (2022).


\bibitem{Huntington} H.B. Huntington, Ch. 6: Electromigration in metals, in: Diffusion in Solids: Recent Developments, Ed. A.S. Nowick, Academic Press, 1975. 

\bibitem{HoKwok} P.S. Ho and T. Kwok, ``Electromigration in metals", \textit{Rep. Prog. Phys.}$\;$ {\bf 52}, 301-348 (1989).

\bibitem{REW} P.J. Rous, T.L. Einstein, and E.D. Williams, ``Theory of surface electromigration on metals: application to self-electromigration on Cu(111)", 
\textit{Surf. Sci.}$\;$ {\bf 315}, L995-L1002 (1994).

\bibitem{SK} M. Schimschak and J. Krug,
``Surface electromigration as a moving boundary value problem",
\textit{Phys. Rev. Lett.}$\;$ {\bf 78}, 278 (1997).



\bibitem{OPLE} O. Pierre-Louis and T.L. Einstein, ``Electromigration of single-layer clusters", \textit{Phys. Rev. B}$\;$ {\bf 62}, 697 (2000). 

\bibitem{KKHV} P. Kuhn, J. Krug, F. Hausser, and A. Voigt, ``Complex Shape Evolution of Electromigration-Driven Single-Layer Islands", 
\textit{Phys. Rev. Lett.}$\;$ {\bf 94}, 166105 (2005).

\bibitem{DDF} M. Dufay, J.-M. Debierre, and T. Frisch,
``Electromigration-induced step meandering on vicinal surfaces: Nonlinear evolution equation",
\textit{Phys. Rev. B}$\;$ {\bf 75},  045413  (2007).

\bibitem{QM} J. Quah and  D. Margetis, ``Electromigration in macroscopic relaxation of stepped surfaces",
\textit{Multiscale Model. and Simul.}$\;$ {\bf 8}, 667 (2010).

\bibitem{SV} D. Solenov and K.A. Velizhanin, ``Adsorbate transport on graphene by electromigration", \textit{Phys. Rev. Lett.}$\;$ {\bf 109}, 095504 (2012).

\bibitem{K} M. Khenner, ``Analysis of a combined influence of substrate wetting and surface electromigration on a thin film stability and dynamical morphologies",
\textit{C. R. Physique}$\;$ {\bf 14}, 607 (2013).

\bibitem{SGM} G.I. Sfyris, M.R. Gungor, and D. Maroudas, ``Analysis of current-driven surface morphological stabilization
of a coherently strained epitaxial thin film on a finite-thickness deformable substrate", \textit{J. Appl. Phys.}$\;$ {\bf 108}, 093517 (2010).

\bibitem{SGM1} G.I. Sfyris, M.R. Gungor, and D. Maroudas, ``Electromigration-driven surface morphological stabilization of a coherently strained epitaxial thin film on a substrate",
\textit{Appl. Phys. Lett.}$\;$ {\bf 96}, 231911 (2010).




\bibitem{DM} L. Du and D. Maroudas,
``Optimization of electrical treatment strategy for surface roughness reduction in conducting thin films",
\textit{J. Appl. Phys.}$\;$ {\bf 124}, 125302 (2018).

\bibitem{CMECPL} S. Curiotto, P. Muller, A. El-Barraj, F. Cheynis, O. Pierre-Louis, and F. Leroy, ``2D nanostructure motion on anisotropic surfaces controlled by electromigration", 
\textit{Appl. Surf. Sci.}$\;$ {\bf 469}, 463-470 (2019).

\bibitem{MySurfSci} M. Khenner, ``Electromigration-guided composition patterns in thin alloy films: a computational
study", \textit{Surf. Sci.}$\;$ {\bf 698}, 121611 (2020).


\bibitem{MyPhysRevMat} M. Khenner, ``Directed long-range transport of a nearly pure component atom clusters by the electromigration of a binary surface alloy", 
\textit{Phys.  Rev. Mater.}$\;$ {\bf 5}, 024001 (2021).  

\bibitem{SMSN} J. Santoki, A. Mukherjee, D. Schneider, and B. Nestler, ``Effect of conductivity on the electromigration-induced morphological evolution of islands with 
high symmetries of surface diffusional anisotropy", \textit{J. Appl. Phys.}$\;$ {\bf 129},  025110 (2021). 

\bibitem{LECMC} 
F. Leroy, A. El Barraj, F. Cheynis, P. Muller, and S. Curiotto, ``Electromigration of Au on Ge(111): Adatom and island dynamics",
\textit{Phys. Rev. B}$\;$ {\bf 106}, 115402 (2022).

\bibitem{CMCL} S. Curiotto, P. Muller, F. Cheynis, F. Leroy, ``Mechanism of droplet motion and in-plane nanowire formation with and without electromigration",
\textit{Appl. Surf. Sci.}$\;$ {\bf 579}, 152015 (2022).




\bibitem{PLAPM} Hongkun Park, Andrew K. L. Lim, A. Paul Alivisatos, J. Park, and Paul L. McEuen, 
``Fabrication of metallic electrodes with nanometer separation by electromigration", 
\textit{Appl. Phys. Lett.}$\;$ {\bf 75}, 301 (1999).

\bibitem{VFDMSKBM} L. Valladares, L.L. Felix, A.B. Dominguez, T. Mitrelias, F. Sfigakis, S.I. Khondaker, C.H.W. Barnes, and Y. Majima,
``Controlled electroplating and electromigration in nickel electrodes for nanogap formation",
\textit{Nanotechnology}$\;$ {\bf 21}, 445304 (2010).

\bibitem{AGLCH} L. Arzubiaga, F. Golmar, R. Liopis, F. Casanova, and L.E. Hueso,
``Tailoring palladium nanocontacts by electromigration",
\textit{Appl. Phys. Lett.}$\;$ {\bf 102}, 193103 (2013).

\bibitem{BUS} D.O. Bellisario, Z. Ulissi, and M.S. Strano,
``A quantitative and predictive model of electromigration-induced breakdown of metal nanowires",
\textit{J. Phys. Chem.}$\;$ {\bf 117}, 12373 (2013).








\bibitem{Korzec} M.D. Korzec, A. M\"{u}nch, and B. Wagner, 
``Anisotropic surface energy formulations and their effect on stability of a growing thin film",
\textit{Interfaces and Free Boundaries}$\;$ {\bf 14}, 545 (2012).

\bibitem{OL} C. Ograin and J. Lowengrub, ``Geometric evolution law for modeling strongly anisotropic thin-film morphology",
\textit{Phys. Rev. E}$\;$ {\bf 84}, 061606 (2011).




\bibitem{TPL} A.K. Tripathi and O. Pierre-Louis, ``Disjoining-pressure-induced acceleration of mass shedding in solid-state dewetting",
\textit{Phys. Rev. E}$\;$ {\bf 101}, 042802 (2020).

\bibitem{KTL} M. Khenner, W.T. Tekalign, and M. Levine, 
``Stability of a strongly anisotropic thin epitaxial film in a wetting interaction with elastic substrate",
\textit{Eur. Phys. Lett.}$\;$ {\bf 93}, 26001 (2011).

\bibitem{K2008} M. Khenner, ``Morphologies and kinetics of a dewetting ultrathin solid film", \textit{Phys. Rev. B}$\;$ {\bf 77}, 245445 (2008).

\bibitem{OPL} O. Pierre-Louis,  ``Solid-state wetting at the nanoscale",
\textit{Progress in Crystal Growth and Characterization of Materials}$\;$ {\bf 62}, 177 (2016).










\bibitem{MSMSE_Khenner} M. Khenner, ``Height transitions, shape evolution, and coarsening of equilibrating quantum nanoislands",
\textit{Modell. Simul. Mater. Sci. Eng.}$\;$ {\bf 25},  085003 (2017).



\bibitem{Mullins95}  W.W. Mullins, ``Mass transport at interfaces in single component systems", \textit{Metall. and Mater. Trans. A}$\;$
{\bf 26}, 1917 (1995).

\bibitem{CT94} J.W. Cahn and J.E. Taylor, ``Surface motion by surface diffusion", \textit{J. Appl. Phys.}$\;$ {\bf 42}, 1045 (1994). 


\bibitem{T} R.S. Timsit, ``Remarks on the thermal stability of an Ohmic-heated nanowire", \textit{J. Appl. Phys.}$\;$ {\bf 123}, 175105 (2018). 


\bibitem{H} R. Hoffmann-Vogel, ``Electromigration and the structure of metallic nanocontacts",
\textit{Appl. Phys. Rev.}$\;$ {\bf 4},  031302 (2017).

\bibitem{XYQF} Luo Xie, Li-jun Yang, Li-zi Qin, Qing-fei Fu, ``Temporal instability of charged viscoelastic liquid jets under an axial
electric field",
\textit{Eur. J. Mech. B/Fluids}$\;$ {\bf 66},  60 (2017).



\bibitem{B1} R.M. Bradley, ``Electromigration-induced propagation of nonlinear surface waves", \textit{Phys. Rev. E}$\;$ {\bf 65}, 036603 (2002).



\bibitem{Herring51} C. Herring, ``Surface tension as a motivation for sintering", Ch. 8, p. 143-179; in: \textit{The Physics of Powder Metallurgy} (1951).


\bibitem{BB} G. Boussinot and E.A. Brener, ``Inhibition of Rayleigh-Plateau instability on a unidirectionally patterned substrate",
\textit{Phys. Rev. E}$\;$ {\bf 92}, 032408 (2015).

 



\end{thebibliography}
\end{document}